\documentclass[twocolumn,showpacs,preprintnumbers,amsmath,amssymb]{revtex4}
\def\address{\affiliation}
\usepackage{graphicx,subfigure}
\newcommand{\parone}[2]{\ensuremath{\frac{\partial #1}{\partial #2}}}
\newcommand{\textparone}[2]{\ensuremath{{\partial #1}/{\partial #2}}}
\newcommand{\bi}{\ensuremath{\mathrm{Bi}_2\mathrm{Sr}_2\mathrm{CaCu}_2\mathrm{O}_{8+\delta}}}
\newcommand{\ent}{\ensuremath{(\partial \mathcal S/ \partial N_e)_{T,V} }}
\newcommand{\eent}{\ensuremath{-(1/e)(\textparone{\mathcal S}{N_e})_{T,V}}}
\begin{document}

\title{
Out-of-plane thermopower of strongly correlated layered systems: an application to Bi$_2$(Sr,La)$_2$CaCu$_2$O$_{8+\delta}$
}

\author{
T. W. Silk$^{1}$, I. Terasaki$^{2,3}$, T. Fujii$^{4}$ and A. J. Schofield$^{1}$}

\address{
$^1$School of Physics and Astronomy, The University of Birmingham, Birmingham, B15 2TT, U.K. \\
$^2$Department of Applied Physics, Waseda University, Tokyo 169-8555, Japan\\
$^3$CREST, Japan Science and Technology Agency, Tokyo 103-0027, Japan\\
$^4$Cryogenic Center, The University of Tokyo, Tokyo 113-0032, Japan 
}

\begin{abstract}

We calculate the out-of-plane thermopower in a quasi-two dimensional system, and argue that this quantity is an effective probe of the asymmetry of the  single-particle spectral function.  We find that the temperature and doping dependence of the out-of-plane thermopower in Bi$_2$(Sr,La)$_2$CaCu$_2$O$_{8+\delta}$ single crystals is broadly consistent with the behavior of the spectral function determined from ARPES and tunneling experiments.  We also investigate the relationship between out-of-plane thermopower and entropy in a quasi-two dimensional material.  We present experimental evidence that at moderate temperatures, there is a qualitative correspondence between the out-of-plane thermopower in Bi$_2$(Sr,La)$_2$CaCu$_2$O$_{8+\delta}$, and the entropy obtained from specific heat measurements.  Finally, we argue that the derivative of the entropy with respect to particle number may be the more appropriate quantity to compare with the thermopower, rather than the entropy per particle.

\end{abstract}

\pacs{74.72.-h, 72.15.Jf, 65.40.gd}

\maketitle

\section{Introduction} \label{sec:Introduction}
As is well known, high-temperature superconductors (HTSC) share the CuO$_2$ plane as a common structural unit, and exhibit highly two-dimensional electronic properties \cite{Ito}.  This extreme anisotropy is demonstrated by the ratio of the in-plane ($a$-$b$-axes) and out-of-plane ($c$-axis) resistivities, $\rho_c/\rho_{ab}$, which can be of the order of $10^3$-$10^5$.  The unusual superconducting and normal-state properties of these materials are thought to originate from strong correlation effects in the $a$-$b$ plane.  Hence, the in-plane charge transport in these materials has been studied much more extensively than the out-of-plane.  However, it has been noted by many authors \cite{kumar_1992a,Nagaosa,Moses,Sandemann} that the out-of-plane transport can be used to provide useful insights about the in-plane physics.  Under certain assumptions, out-of-plane transport properties depend only on the in-plane single-particle spectral function, and the out-of-plane hopping integral.  To date, attention has primarily focused on the out-of-plane resistivity and magnetoresistance, with much less attention given to thermal and thermoelectric properties.   

The significance of thermopower has, however been recognized in recent years.  First, it is a quantity independent of sample dimension, and, like the Hall effect, is often insensitive to the grain boundary and/or disorder.  Second, most of the transition-metal oxides rarely show complicated non-equilibrium phenomena such as phonon drag, because of poor mobility.  Recent work by Kontani \cite{kontani} has suggested that the in-plane thermopower of HTSC can be rather simply understood within spin-fluctuation theory. 

The electronic contribution to the thermopower of a material is fundamentally related to the \textit{difference} in the response of electrons and holes to an applied temperature gradient \cite{Ashcroft}.  The thermopower therefore probes the particle-hole asymmetry of the system.  The $c$-axis thermopower is particularly useful in this regard, since it can be directly related to the particle-hole asymmetry of the single-particle spectral function. 

In the first part of this paper we will give an explicit calculation of the $c$-axis thermopower within the tunneling Hamiltonian formalism, and show that the result can be written as a Mott formula.  Using recent experimental measurements of the $c$-axis thermopower of $\bi$ (Bi2212) single crystals, we then investigate what can be inferred about the asymmetry of the single particle spectral function around the chemical potential.  Together with the $c$-axis resistivity, which primarily reflects the symmetric part of the spectral function, these two quantities suggest that the spectral function becomes weakly temperature dependent at high temperatures.  At lower temperatures, the $c$-axis thermopower may reflect both the physics of the pseudogap, and superconducting fluctuations.  We show that the properties of the spectral function inferred from $c$-axis transport are in broad agreement with information provided from ARPES and tunneling spectroscopy. 

We note that the \textit{opposite} procedure, starting with the electronic structure, is often used to investigate the thermopower.  Within a Boltzmann transport picture, and under the assumption that the relaxation time is an energy independent constant, the thermopower is determined entirely by the band structure of the material.  This approach has been used, for example, to calculate the in-plane thermopower of the layered cobalt oxide Na$_x$Co$_2$O$_4$. The necessary quantities are either obtained from electronic structure calculations \cite{Singh}, or by using photoemission spectroscopy to determine the density of states, combined with some assumptions about the energy dependence of the group velocity \cite{Takeuchi}.       

The second part of this paper explores the relationship between the $c$-axis thermopower and the entropy of the system.  The basis for this relationship are the Kelvin/Onsager relations \cite{deGroot,Heikes}, which relate the Seebeck and Peltier coefficients of a material.  This will be discussed further in section \ref{sec:Results}: here, we will use an empirical relationship recently noted by Behnia \textit{et al.} \cite{Behnia} as motivation.  For non-interacting electrons with a quadratic dispersion and an energy-independent relaxation time, it can be shown that the thermopower $S$ \textit{is} just the equilibrium entropy $\mathcal S$ per electron, $S = -\mathcal S/N_ee$.  Behnia \textit{et al} have considered the quantity $q=eS/C$, where $C$ is the electronic specific heat per electron/hole (for non-interacting electrons at low temperatures, $C$ is also equal to the entropy per charged carrier).  The quantity $q$ is equal to $-1$ for free electrons, and $+1$ for free holes.  Empirically, it was noted that if $q$ is calculated for a range of correlated materials from the low temperature (\textit{i.e.} $T \rightarrow 0$) thermopower and specific heat, values of $q$ were still typically close to $\pm 1$.  Miyake and Kohno \cite{Miyake} have shown that this universal value of $q$ in the $T \rightarrow 0$ limit can be understood within a Fermi-liquid picture.  Zlati\'{c} \textit{et al} \cite{Zlatic} have also investigated this relationship in Ce, Eu and Yb-based heavy Fermion compounds via a DMFT approach.
 
Another limit in which the thermopower is directly related to the equilibrium entropy is in finite bandwidth systems at temperatures much larger than the kinetic energy scale.  In this limit the thermopower is related to the derivative of the entropy with respect to particle number, rather than to the entropy per particle.  The resulting expressions are referred to as Heike's formulae \cite{Beni,Chaikin}.  

Despite the existence of `exact' results in the particular cases indicated above, it is clear that the relationship between thermopower and equilibrium entropy cannot be truly universal.  This fact is obviously demonstrated by the existence of materials like Bi2212 where the thermopower is anisotropic.  If the thermopower in one direction is closely related to the equilibrium entropy, then the thermopower in other directions will not be. Nevertheless, given the general difficulties with interpreting the thermopower, any relationships which can be established with more familiar quantities are very useful.  It is therefore important to identify those special cases where relationships with the entropy hold, and to understand the reasons for this. 

In this paper we will discuss the relationship between thermopower and entropy in general terms, using the thermopower (calculated in the relaxation time approximation) and entropy of electrons in a 2d tight-binding band as a simple example.  In particular, we argue that the derivative of the entropy with respect to particle number may be the more fundamental quantity to compare with the thermopower, something which has not been previously recognized.  We argue that the $c$-axis thermopower of a quasi-two dimensional material is one of the special cases where a close relationship with the equilibrium entropy might be expected to hold.  We will see experimentally that at moderate temperatures there is a qualitative correspondence with the entropy per particle, and that theoretically a similar qualitative correspondence can be established with the derivative of the entropy with respect to particle number. 

The rest of this paper is organized as follows.  In Sec.~\ref{sec:Experimental} we analyze experimental measurements of the $c$-axis thermopower in Bi2212 single crystals from Ref.~\cite{TerasakiFujii}.  The calculation of the $c$-axis thermopower is given in Sec.~\ref{sec:Theory}.  In Sec.~\ref{sec:Bi2212}, we discuss the form of the tunneling matrix element in Bi2212, and compare the information about the asymmetry of the spectral function provided by the $c$-axis thermopower with that obtained from ARPES and tunneling spectroscopy.  In Sec.~\ref{sec:Results}, we discuss in general terms the relationship between thermopower and entropy, using the non-interacting 2d tight-binding model as an example.  Finally, in Sec.~\ref{sec:caxis} we compare the experimentally measured thermopower with the entropy from specific heat measurements, and theoretically compare the $c$-axis thermopower with the derivative of the entropy with respect to particle number.

\section{c-axis thermopower of Bi2212}\label{sec:Experimental}

\begin{figure}[t]
 \begin{center}
  \includegraphics[width=0.5\textwidth]{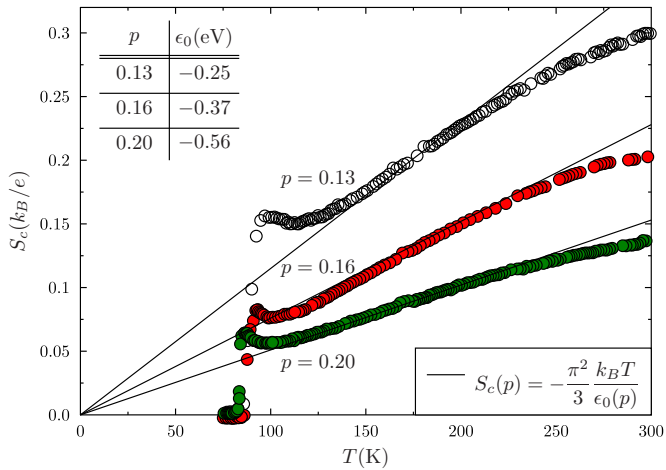}
 \end{center}
 \caption{\label{fig:fits} Out-of-plane thermopower of Bi2212 single crystals (reported in Ref.~\cite{TerasakiFujii}), for three different doping levels, and in the natural unit of $k_B/e$.  The doping level is quantified by $p$, the carrier concentration per Cu.  Also shown are fits to the thermopower over the appropriate temperature range (see main text).  The inset shows the fit parameters for each doping.}
\end{figure}

Here we consider the $c$-axis thermopower data from Bi2212 single crystals reported in Ref.~\cite{TerasakiFujii} that provides motivation for this work.  Fig.~\ref{fig:fits}(a) shows the temperature dependence of the $c$-axis thermopower ($S_c$) in Bi2212 with various oxygen contents.  The respective values of $T_c$ for $p=0.13$, 0.16 and 0.20 are 85K, 89K and 84K.  The carrier concentration per Cu ($p$) is estimated from the empirical relation to the room-temperature in-plane thermopower $S_{ab}$ \cite{tallon}.  Unlike $S_{ab}$ \cite{TerasakiFujii}, the sign of $S_c$ is positive over the measured temperature range for all dopings.  The key features of the data are i) an approximately linear temperature dependence at moderate temperatures, becoming sublinear at higher temperatures; ii) the slope of the linear part is positive, and has a magnitude which decreases with doping; iii) an upturn in the thermopower at lower temperatures, with the temperature at which the upturn occurs decreasing with doping.   

Having identified the salient features of the data, we now turn to the theoretical calculation of the $c$-axis thermopower.  

\section{Theory of c-axis Thermopower}\label{sec:Theory}

We consider the $c$-axis thermopower of a quasi-two dimensional material consisting of well-separated planes of atoms in which electrons move relatively easily, but interplane motion is suppressed.  This has two important consequences.  First, the interplane hopping integral, $t_c$ is much smaller than the typical in-plane hopping integral $t_{ab}$.  Second, interactions between electrons in neighboring planes are expected to be weaker than between electrons in the same plane.  Further, we will assume that each layer is translationally invariant, and hence the in-plane momentum is conserved in the tunneling process.  This assumption could be violated if, for example, the $c$-axis transport occurred via the resonant tunneling of the electron through impurities located between the layers \cite{maslov_2008a}.   

The most suitable formalism in which to calculate the $c$-axis thermopower is the tunneling Hamiltonian \cite{Mahan}.  In this approach, transport quantities are calculated perturbatively, with the small parameter taken to be $t_c$, rather than the magnitude of the applied field.  Further, if interactions between the layers are neglected, then there are no vertex corrections to the elementary bubble.  We stress that in using the tunneling Hamiltonian approach with $t_c$ treated as a small quantity, we have not assumed that the $c$-axis transport is necessarily incoherent \cite{Moses}.  The coherence issue depends on the relative size of the Fermi energy $\epsilon_F$ and the inelastic scattering time $\tau$ \cite{Maslov}.  Since $\epsilon_F$ will depend on both $t_{ab}$ and $t_c$, it is completely possible to have coherent transport with $\epsilon_F \tau \gg \hbar$, despite having $t_c\tau \ll \hbar$.                 

We start with the Hamiltonian of the complete system (\textit{i.e.} all layers) in a real space representation, and then transform into momentum space along the in-plane directions only.  The Hamiltonian is then written in terms of operators $a^{\dagger}_{\mathbf k,i}$ (we have suppressed spin indices here), which create an electron in the $i$th layer with in-plane momentum $\mathbf k$.  We assume translational invariance in the in-plane direction, and hence the tunneling matrix element satisfies $T_{\mathbf{pk}} = t_{\mathbf k}\delta_{\mathbf p \mathbf k}$.  The form of $t_{\mathbf k}$ is arbitrary at this stage.  We then separate out the part of the kinetic energy which hops particles in the $c$-axis direction \textit{i.e.} transfers particles between layers.  We further assume that this kinetic term is the \textit{only} term in the Hamiltonian which couples one layer to the other.  This leads us to the following Hamiltonian
\begin{equation}
H = \sum_i H_i + \sum_{i,\mathbf k} t_{\mathbf k} a^{\dagger}_{\mathbf k,i+1}a_{\mathbf k,i} + \mathrm{h.c.},
\end{equation}  
where $H_i$ refers to the in-plane Hamiltonian, which contains only operators belonging to the $i$th layer.  

To convert this into a tunneling problem, we argue that to lowest order in $t_{\mathbf k}$, the current of electrons along the $c$-axis is limited by the current passing between just two planes.  We can then simply isolate \textit{any} two planes, and use the tunneling approach to calculate the current to lowest order in $t_{\mathbf k}$.  We therefore work with the Hamiltonian 
\begin{align}
H' &= H_i + H_{i+1} + \sum_{\mathbf k} t_{\mathbf k} a^{\dagger}_{\mathbf k,i+1}a_{\mathbf k,i} + \mathrm{h.c.} = H_0 + H_T.
\end{align}
Since the Hamiltonians $H_i$ and $H_{i+1}$ contain only operators in the $i$th and $(i+1)$th layers respectively, they commute with each other and effectively represent independent systems. 
  
To calculate the thermopower in the tunneling Hamiltonian formalism, we exploit the Kelvin relation between the thermopower and Peltier coefficient, $\Pi = TS$.  Recall that $\Pi$ is the coefficient of proportionality between the heat current and the electrical current flowing under isothermal conditions.  The total heat current is related to the energy current by $Q = U + (\mu/e) I$.  We therefore use the tunneling Hamiltonian approach to calculate the energy and electrical currents between the layers in response to an applied potential difference $V$.  The electrical and energy currents are given by  
\begin{align}
I(t) &= -\frac{e}{2}\langle (\dot N_i - \dot N_{i+1})\rangle  = -\frac{ie}{2}\langle[H_T,N_{i}-N_{i+1}]\rangle \nonumber \\
&= ie\sum_{\mathbf k}(t_{\mathbf{k}}a^{\dagger}_{\mathbf k_{||},i+1}a_{\mathbf k_{||},i} - \mathrm{h.c.}), \\
U(t) &= \frac{1}{2}\langle (\dot H_i - \dot H_{i+1}) \rangle = \frac{i}{2}\langle[H_T,H_i-H_{i+1}]\rangle.
\end{align}
Unlike the electrical current, an explicit form cannot be obtained for the energy current, since we have not specified the form of the in-plane Hamiltonian.  However, within the tunneling Hamiltonian approach this does not matter, since we can obtain a \textit{general} expression for the energy current without specifying the in-plane Hamiltonian explicitly.
\begin{figure}[t]
  \centering
  \subfigure[]{\includegraphics[width=0.23\textwidth]{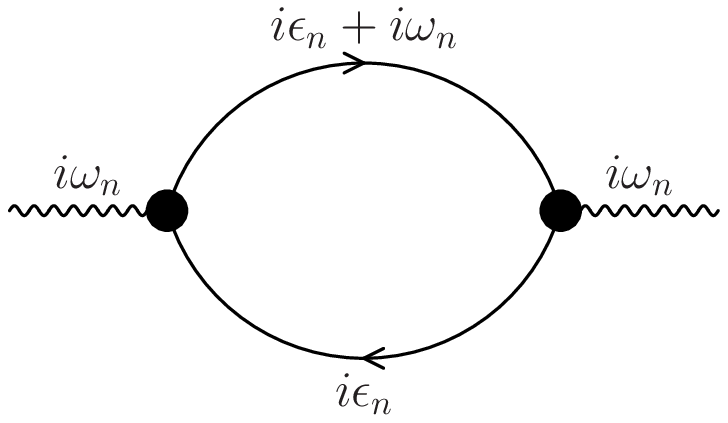}}       
  \subfigure[]{\includegraphics[width=0.23\textwidth]{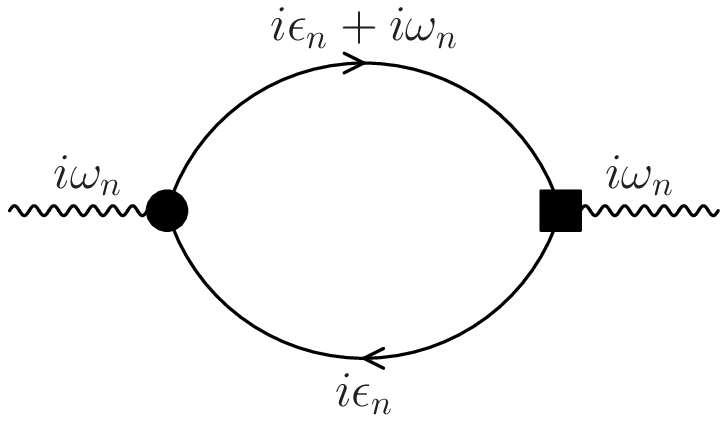}} 
  \caption{The diagrams evaluated in the calculation of the thermopower, where $i\epsilon_n$ and $i\omega_n$ are the fermionic and bosonic Matsubara frequencies respectively.  Circles indicate vertices proportional to $t_{\mathbf k}$, while squares indicate vertices proportional to $t_{\mathbf{k}}(i\epsilon_n + i\omega_n/2)$.  (a) $\Pi(i\omega_n)$. (b) $\Omega(i\omega_n)$.}
  \label{fig:bubblefig}
\end{figure}

The electrical and energy currents are calculated using the Kubo formula with $H_T$ treated as the perturbation.  We find, to lowest order in $t_{\mathbf k}$: 
\begin{align}
&I(V) = -2e\mathrm{Im}[\Pi_{ret}(-eV)], \\
&\displaystyle{U(V) = 2\mathrm{Im}[\Omega_{ret}(-eV)] - \frac{(2\mu+eV)}{2e}I}, \\
& \displaystyle{\Pi(i\omega_n) =  \sum_{\mathbf k} |t_{\mathbf{k}}|^2 \frac{1}{\beta}\sum_{i\epsilon_n}\mathcal G(\mathbf k,i\epsilon_n) \mathcal G(\mathbf k,i\epsilon_n + i\omega_n)},
\end{align}
\vspace{-0.7cm}
\begin{multline}
 \Omega(i\omega_n) =
 \sum_{\mathbf k} |t_{\mathbf{k}}|^2 \frac{1}{\beta}\sum_{i\epsilon_n}(i\epsilon_n + i\omega_n/2) \\
\times \mathcal G(\mathbf k,i\epsilon_n) \mathcal G(\mathbf k,i\epsilon_n + i\omega_n).
\end{multline}
These correlation functions are shown diagrammatically in Fig.~\ref{fig:bubblefig}.  The absence of interactions between the layers means that there are no vertex corrections to these elementary bubbles. The thermopower is obtained via the Kelvin relation as $S_c = \lim_{V \rightarrow 0} (1/T)Q(V)/I(V)$: 
\begin{align}
\label{eq:caxisthermo}
S_c = -\frac{1}{e}\frac{{\displaystyle\sum_{\mathbf k} t_{\mathbf{k}}^2\int d \epsilon 
A^2({\mathbf k},\epsilon)\frac{\epsilon-\mu}{T} \Big(-\frac{\partial f}{\partial 
\epsilon}\Big)}}{\displaystyle{\sum_{\mathbf k} t_{\mathbf{k}}^2 \int d \epsilon A^2({\mathbf k},\epsilon) \Big(-\frac{\partial f}{\partial \epsilon}\Big)}},
\end{align}
where $A(\mathbf k, \epsilon) = -(1/\pi)\mathrm{Im}[G^R(\mathbf k,\epsilon)]$ is the single-particle spectral function, $f(\epsilon) = (1+e^{\beta(\epsilon-\mu)})^{-1}$ is the Fermi function, and $\mu$ is the chemical potential.  This result coincides with that found by Nagaosa \cite{Nagaosa}.  If we define an energy-dependent conductivity  $\sigma_c(\epsilon) = 2e^2\sum_{\mathbf k} t_{\mathbf k}^2 A^2(\mathbf k,\epsilon)$, then the $c$-axis thermopower may be written as 
\begin{equation}
S_c = -\frac{1}{e}\frac{\displaystyle{\int d\epsilon \sigma_c(\epsilon) \Big(\frac{\epsilon-\mu}{T}\Big)\Big(-\parone{f}{\epsilon}\Big)}}{\displaystyle{\int d\epsilon \sigma_c(\epsilon) \Big(-\parone{f}{\epsilon}\Big)}},
\end{equation}
which is known as a Mott formula.  Note that this result has been derived for an \textit{arbitrary} in-plane spectral function, and no assumptions have been made about the nature of the in-plane physics.  We have shown explicitly that within the tunneling Hamiltonian approach, the thermopower may always be expressed as a Mott formula \cite{Mottform}.

Eq.~(\ref{eq:caxisthermo}) shows that the $c$-axis thermopower probes the asymmetry of the spectral function around the Fermi energy.  To demonstrate this explicitly, we write the spectral function in terms of symmetric and antisymmetric parts: $A(\mathbf k, \epsilon) = A_s(\mathbf k, \epsilon) + A_a(\mathbf k, \epsilon)$, where $A_s(\mathbf k,\omega) = A_s(\mathbf k,-\omega)$, $A_a(\mathbf k,\omega) = -A_a(\mathbf k,-\omega)$ and $\omega = \epsilon - \mu$.  Using these definitions, the $c$-axis thermopower may be written as
\begin{align}
\label{eq:caxisthermosa}
S_c = -\frac{1}{e}\frac{{\displaystyle\sum_{\mathbf k} t_{\mathbf{k}}^2\int d \epsilon 
\big ( 2A_a({\mathbf k},\epsilon)A_s({\mathbf k},\epsilon)\big )\frac{\epsilon-\mu}{T} \Big(-\frac{\partial f}{\partial 
\epsilon}\Big)}}{\displaystyle{\sum_{\mathbf k} t_{\mathbf{k}}^2 \int d \epsilon \big( A_s^2({\mathbf k},\epsilon)+A_a^2({\mathbf k},\epsilon)\big ) \Big(-\frac{\partial f}{\partial \epsilon}\Big)}}.
\end{align}
In particular, if the spectral function is particle-hole symmetric for all $k$ (\textit{i.e.} $A_a(\mathbf k, \epsilon)=0$), the numerator of (\ref{eq:caxisthermosa}) vanishes, and the $c$-axis thermopower is zero.  Note that the $c$-axis conductivity, which is (up to a constant of proportionality) the denominator of Eq.~(\ref{eq:caxisthermosa}), also probes both the symmetric and antisymmetric parts of the spectral function.  However, because $A(\mathbf k, \epsilon) \ge 0$, and hence $|A_s(\mathbf k, \epsilon)| \ge |A_a(\mathbf k, \epsilon)|$, it follows that the conductivity is dominated by the symmetric part of the spectral function.    

In the next section, we will apply these ideas to Bi2212.  We will investigate what information about the symmetry properties of the spectral function can be inferred from the $c$-axis thermopower, and whether this is consistent with ARPES and tunneling spectroscopy.  

\section{Application to B$\textrm i$2212}\label{sec:Bi2212}

Each momentum state in the energy-dependent conductivity is weighted by the matrix element $t_{\mathbf k}$, and it is important to understand the effect that this has. In cuprates with a simple tetragonal structure and one CuO$_2$ plane per unit cell, it is well known that the tunneling matrix element depends strongly on the in-plane momentum, and vanishes along the zone diagonals \cite{Andersen}.  The origin of this effect is that the hopping process occurs via the Cu $4s$ orbital.  The overlap of this orbital with the in-plane Cu-O hybrid orbitals has $d_{x^2-y^2}$ symmetry.  Because the $c$-axis tunneling matrix element vanishes along certain directions in $k$-space, one of its most important roles in $c$-axis transport is to select the region of $k$-space which makes the dominant contribution \cite{Xiang,Millis}. 

For Bi2212, two additional complications arise.  The first is that the underlying lattice in this material is body-centered tetragonal.  The off-set this creates between neighboring CuO$_2$ planes means that the hopping matrix element now vanishes along the zone-boundaries as well as along the zone-diagonals \cite{vandermarel}.  

The second complication is that Bi2212 is a bilayer compound, with two CuO$_2$ planes per unit cell.  A minimal description necessarily involves two hopping parameters.  One describes the hopping within the bilayer (and is responsible for the bilayer splitting), and the other involves hopping between different bilayers.  In the limit where the bilayer hopping $t_{bi}$ is much larger than the inter-bilayer hopping $t_c$, the bilayer splitting can be separated from the intercell hopping, giving the following expression for the tunneling matrix element \cite{Markiewicz}:
\begin{multline}\label{eq:tunmatel}      
t_{\mathbf k} = \pm 4t_c\cos(k_xa/2)\cos(k_ya/2)\times \\
\big( (\cos k_xa - \cos k_ya)^2/4 + a_0 \big ),
\end{multline}
where $a$ is the in-plane lattice spacing.  The `vertical hopping' parameter $a_0$ has been introduced phenomenologically in Ref.~\cite{Markiewicz} to account for hopping processes which are not assisted by Cu 4$s$ orbitals. It is needed to fit the LDA band-structure, where a finite bilayer splitting is seen at the zone-center.  Note that thermopower does not depend directly on the magnitude of $t_c$, since it cancels from the numerator and denominator of Eq.~(\ref{eq:caxisthermo}).  

The spectral function in Bi2212 has been extremely well characterized by ARPES \cite{Damascelli}.  However, it is important to remember that the ARPES intensity $I(\mathbf k,\omega)$ is proportional to $f(\omega)A(\mathbf k, \omega)$, and so only the occupied states are probed.  Norman \textit{et al} \cite{Norman, Norman2} have proposed a technique to eliminate the effects of $f(\omega)$ from ARPES data.  Under the assumption that the spectral function is particle-hole symmetric at low energies and momenta close to the Fermi wavevector, the symmetrized intensity $I(\omega) + I(-\omega)$ is just the spectral function (convolved with the resolution).  However, since the $c$-axis thermopower is non-zero (\textit{cf} Fig.~\ref{fig:fits}), we know that this assumption of particle-hole symmetry cannot be strictly correct.  We can use the $c$-axis thermopower to obtain a measure of this asymmetry and to indicate whether important physics is being lost by an analysis that assumes the spectral function to be symmetric. 

Let us consider the high temperature regime, \textit{i.e.} at temperatures well above the pseudogap onset temperature $T^*$ \cite{Timusk}.  The temperature $T^*$ decreases with doping, and is eventually cut-off by $T_c$ in the overdoped regime.  Above $T^*$, ARPES suggests that spectral function for the occupied states becomes only weakly dependent on temperature \cite{Norman}, and hence the temperature dependence of the thermopower is dominated by the Fermi functions.  Assuming that the energy-dependent conductivity varies slowly on the scale set by $k_B T$, the Sommerfeld approximation may be used to evaluate the thermopower:  
\begin{align}\label{eq:Somm}
S_c &= -\frac{\pi^2}{3}\frac{k_B^2T}{e}\frac{\sigma_c '(\epsilon _F)}{\sigma_c (\epsilon_F)} = -\frac{\pi^2}{3}\frac{k_B}{e}\frac{k_BT}{\epsilon_0}.
\end{align}
The thermopower in this regime is therefore expected to be linear in temperature.  In Fig.~\ref{fig:fits} we show the results of fitting this form to the $c$-axis thermopower in the appropriate temperature range.  The sublinear behavior at very high temperatures has two possible causes.  One possibility is that the spectral function has some stronger temperature dependence not seen at lower temperatures.  A second possibility is that at higher temperatures the Fermi functions explore $\sigma_c(\epsilon)$ over a greater energy range.  This may invalidate the Sommerfeld approximation, although the exact temperature dependence which results will depend sensitively on the exact shape of $\sigma_c(\epsilon)$ \cite{Kaiser}.

The quantity $\sigma_c(\epsilon_F)/\sigma_c'(\epsilon_F)$ defines an energy scale $\epsilon_0$.  For free electrons in $d$-dimensions where the energy dependent conductivity varies as a power-law in energy (see Sec.~\ref{sec:Results}), this energy scale is proportional to the Fermi energy: $\epsilon_0 = 2\epsilon_F/d$.  The extracted values of the energy scale $\epsilon_0$ are shown in the inset to Fig.~\ref{fig:fits}.  The energy scale increases with the number of carriers $p$, while the sign reflects the sign of the charge carriers (holes).

We can obtain an independent measure of the expected scale of $\sigma_c'/\sigma_c$ from tunneling experiments on this material \cite{Renner}.  The tunneling conductance from an STM tip is related to the local density of states, $\nu(\epsilon) = 2\sum_{\mathbf k}A(\mathbf k,\epsilon)$.  At high temperatures, the spectrum is fairly smooth and temperature-independent, but is not constant in energy.  If we use the data of Ref.~\cite{Renner} to estimate the value of the quantity $\nu/\nu'$ over the band, we find a value of $-0.68$eV for an underdoped ($T_c=83$K) sample, and $-0.84$eV for an overdoped ($T_c=74.3$K) sample.  These energy scales are of the same order of magnitude, and have the same sign as those extracted from the $c$-axis thermopower.

Further support for the assertion that the spectral function becomes only weakly temperature-dependent above $T^*$ comes from the $c$-axis resistivity.  Using the Sommerfeld approximation, we would expect that in this temperature regime the $c$-axis resistivity should be approximately independent of temperature.  This is in good agreement with experiment \cite{fujii2, Watanabe2}, where the $c$-axis resistivity becomes only weakly temperature dependent at high temperatures. 

We now turn to the behavior of the thermopower at lower temperatures.  Here, the thermopower is seen to deviate from linearity, pass through a maximum, and then finally fall to zero below $T_c$.  Two effects become important in this regime.  The first is the effect of the pseudogap, particularly in the underdoped  and optimally doped samples.  The effect of the pseudogap on the energy dependent conductivity is modulated by the tunneling matrix element, which selects the regions of $k$-space which make the dominant contribution.  Because $t_{\mathbf k}$ vanishes along the zone-boundaries in Bi2212, the region around $(\pi,0)$ point where the pseudogap is largest does not contribute significantly.  Nevertheless, we would still expect to see some suppression of the energy-dependent conductivity around the Fermi energy, although possibly to a lesser extent than that observed in the density of states.  

This suppression of the energy-dependent conductivity has been linked to the increase in the $c$-axis resistivity below $T^*$ \cite{Zha, Loram2}.  Since the conductivity appears as the denominator in the expression for the thermopower, it is tempting to associate the decrease in the conductivity with the increase in the thermopower.  The temperature at which the upturn in the thermopower occurs decreases with doping, which is consistent with the behavior of the pseudogap opening temperature $T^*$.  This argument neglects the possible temperature dependence in the numerator of (\ref{eq:caxisthermo}), which will more strongly reflect the changing asymmetry of the spectral function in the pseudogap regime.  It would be interesting to combine both $c$-axis thermopower and resistivity data together to form the quantity $\sigma_c S_c$.  This would indicate the extent to which the upturn in the thermopower is due to changes in the asymmetry of the spectral function. However $\sigma_c$ has not been published for this crystal. 

Very close to $T_c$, a second possible contribution to the upturn in the thermopower are superconducting fluctuations.  To our knowledge, the fluctuation contribution to the $c$-axis thermopower has not been calculated previously.  Ioffe \textit{et al.} \cite{Ioffe2} have calculated the fluctuation contribution to the $c$-axis conductivity.  Near to $T_c$, they have identified a contribution which leads to a \textit{decrease} in the conductivity, which is subsequently suppressed by the usual divergence at $T_c$.  This decrease in the conductivity suggests a possible explanation for the sharp rise in the $c$-axis thermopower close to $T_c$.  This argument is particularly relevant for the $p=0.20$ sample.  The pseudogap, as determined from specific heat measurements \cite{loram}, is expected to vanish around $p=0.19$, and hence may not explain the upturn in the thermopower in the $p=0.20$ sample.  
\begin{figure}[t]
 \begin{center}
  \includegraphics[width=8cm,clip]{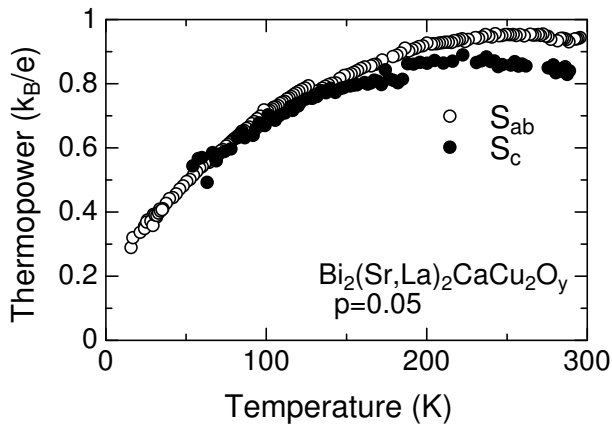}
 \end{center}
 \caption{\label{fig:insulator}
 In-plane and out-of-plane thermopower of single crystals of the parent compound
 Bi$_2$(Sr,La)$_2$CaCu$_2$O$_8$ (after Ref.~\cite{TerasakiFujii}).  
 }
\end{figure}

For completeness, we conclude our discussion of Bi2212 by considering the $c$-axis thermopower of the undoped parent compound.  Fig.~\ref{fig:insulator} shows $S_c$ and $S_{ab}$ for the parent ($T=0$) insulator Bi$_2$(Sr,La)$_2$CaCu$_2$O$_8$, reproduced from Ref.~\cite{TerasakiFujii}.  Remarkably, the thermopower of the parent compound is nearly isotropic, which suggests that the mechanism of transport in the insulating state is the same in- and out-of-plane.  It has been proposed that the in-plane resistivity and thermopower \cite{takemura} can be explained in terms of three-dimensional variable-range hopping \cite{Burns}.  This approach gives $S_{ab} \sim \sqrt{T}$ at low temperatures, and hence one may infer from the data that $S_c \sim \sqrt{T}$ also.  Under the assumption that our theory of $c$-axis thermopower still applies to the parent compound, this result would provide information on the temperature dependence of the spectral function.  However, this assumption may be invalidated by the fact that the parent compound is magnetically ordered, and so correlations between layers may have a significant effect on $c$-axis transport. 

In summary, by comparing our analysis of the $c$-axis thermopower to published experimental data on Bi2212, we can draw the following conclusions.  By assuming that the spectral function becomes independent of temperature above $T^*$, we have shown that the $c$-axis thermopower is expected to be linear in temperature and the $c$-axis resistivity temperature independent, at least while the Sommerfeld approximation remains valid.  Comparison with the experimental data has confirmed these predictions, and so we conclude that both the symmetric and antisymmetric parts of spectral function become weakly temperature dependent above $T^*$.  The antisymmetric part of the spectral function is characterized by energy scales consistent with those seen in tunneling.  Just above $T_c$, the thermopower shows an upturn whose microscopic origin could be related to the pseudogap physics and/or superconducting fluctuations.  A comparison with resistivity measurements on the same crystals would reveal if this were due to changes in the antisymmetric part of the spectral function.

\section{Relation of Thermopower to Entropy}\label{sec:Results}

In this section we will discuss the relationship between thermopower and entropy in general terms.  Recall that the thermopower or Seebeck coefficient of a sample is defined as the ratio of the induced voltage gradient to an applied temperature gradient under conditions of zero electrical current flow, $S = (dV/dT)_{j=0}$.  Under isothermal conditions the heat current density, $j_Q$, associated with a given electrical current density, $j$, is given by $j_Q=\Pi j$, where $\Pi$ is the Peltier coefficient.   The Kelvin/Onsager relations \cite{deGroot,Heikes} show that the Seebeck and Peltier coefficients are related by $\Pi = ST$.  It follows that we can write the thermopower as
\begin{equation}\label{eq:entthermo}
S = \Pi/T = (j_{\mathcal S}/j)_{\nabla T=0},
\end{equation}
where $j_{\mathcal S} = j_Q/T$ is the entropy current density.  The question is then how to interpret final equality on the right-hand side of Eq.~(\ref{eq:entthermo})?  The conventional interpretation is to say that \textit{``the thermopower is the transported entropy per charge carrier''}.  However, given that a current density is a \textit{rate} of flow, an alternative interpretation of the thermopower is \textit{``the rate of change of transported entropy with number of charge carriers''}.  We also need to define the terms `transported entropy' and `charge carrier': is the transported entropy the same as the equilibrium entropy, and what is the relevant number of charge carriers?  

To further explore these issues, let us examine the case of non-interacting electrons in a single band, with an arbitrary dispersion relation.  The thermopower is calculated from the Boltzmann equation in the relaxation time approximation  \cite{Ashcroft}.  If the system is anisotropic, then the thermopower will generally be a 2nd rank tensor quantity: here, we choose to focus on just one of the diagonal components, $S_x$.  In addition, we consider the quantities $-\mathcal S/N_ee$, $\mathcal S/N_he$ and $-(1/e)(\textparone{\mathcal S}{N_e})_{T,V}$, where $N_e$ is the number of electrons and $N_h$ is the number of holes.  Straightforward calculations provide:
\begin{equation}\label{eq:elecent}
\frac{\mathcal S}{N_e} =-k_B\frac{\displaystyle{\int d\epsilon \nu(\epsilon)\big[f\ln f + (1-f)\ln(1-f) \big]}}{\displaystyle{\int d\epsilon \nu(\epsilon)f}},
\end{equation} 
\begin{equation}\label{eq:holeent}
\frac{\mathcal S}{N_h} =-k_B\frac{\displaystyle{\int d\epsilon \nu(\epsilon)\big[f\ln f + (1-f)\ln(1-f) \big]}}{\displaystyle{\int d\epsilon \nu(\epsilon)(1-f)}},
\end{equation}
\begin{eqnarray}
-eS_x = F[\sigma_x(\epsilon)], & \displaystyle{\Big (\parone{\mathcal S}{N_e}\Big )_{T,V} = F[\nu(\epsilon)]},
\end{eqnarray}
where
\begin{equation}\label{eq:Afunc}
F[g(\epsilon)] = \frac{\displaystyle{\int d\epsilon g(\epsilon) \Big(\frac{\epsilon-\mu}{T}\Big)\Big(-\parone{f}{\epsilon}\Big)}}{\displaystyle{\int d\epsilon g(\epsilon) \Big(-\parone{f}{\epsilon}\Big)}},
\end{equation}
and where $\sigma_x(\epsilon) = 2e^2\sum_{\mathbf k}v_x^2(\mathbf k)\tau(\epsilon(\mathbf k))\delta(\epsilon-\epsilon(\mathbf k))$ is the energy-dependent conductivity and $\nu(\epsilon) = 2\sum_{\mathbf k}\delta(\epsilon-\epsilon(\mathbf k))$ is the density of states.  These expressions reveal a fundamental difference between $\mathcal S/N_e$ and $\mathcal S/N_h$ on the one hand, and $S_x$ and $\ent$ on the other.  Both $\mathcal S/N_e$ and $\mathcal S/N_h$ can be thought of as the \textit{sum} of (positive) contributions from electrons and holes.  This means that $\mathcal S/N_e$ and $\mathcal S/N_h$ are always positive quantities. In contrast, $S_x$ and $\ent$ should be thought of as the \textit{difference} of contributions from electrons and holes.  These quantities can be either positive if holes dominate, negative if electrons dominate, or zero if the system is particle-hole symmetric.

It is also possible to express $\mathcal S/N_e$ and $\mathcal S/N_h$ using the functional $F$.  Integrating by parts in the numerator and denominator of (\ref{eq:elecent}) and (\ref{eq:holeent}), we find that $\mathcal S/N_e = F[\Gamma(\epsilon)]$ and $\mathcal S/N_h = F[N_s-\Gamma(\epsilon)]$, where $\Gamma(\epsilon) = \int_{-\infty}^{\epsilon} \nu(\epsilon')d\epsilon'$ is the energy-dependent electron number, and $N_s = \Gamma(\infty)$ is the total number of states.   

Assuming that the function $g(\epsilon)$ varies slowly on a scale set by $k_B T$ around $\epsilon=\mu\approx \epsilon_F$, the Sommerfeld approximation may used to evaluate the integrals in (\ref{eq:Afunc}), giving
\begin{align}
F[g(\epsilon)] \approx \frac{\pi^2}{3}k_B^2T\frac{g'(\epsilon_F)}{g(\epsilon_F)},
\end{align}
where $g' \equiv \textparone{g}{\epsilon}$.  Hence $F[g]$ is expected to be linear in temperature while the Sommerfeld approximation holds.  The functional $F[g]$ measures the relative asymmetry of the function $g(\epsilon)$ about the Fermi energy.  The thermopower and $\ent$ can either be positive or negative, depending on the signs of $\sigma_x'(\epsilon_F)$ and $\nu'(\epsilon_F)$.  The positive nature of $\mathcal S/N_e$ and $\mathcal S/N_h$ follows since $\Gamma'(\epsilon) = \nu(\epsilon) \ge 0$.   

Let us now consider the two limits described in the introduction, where exact relationships between the thermopower and entropy hold.  For non-interacting electrons with a quadratic dispersion in $d$-dimensions, it can be shown that $\sigma_x(\epsilon) \propto \tau(\epsilon)\epsilon^{d/2}$ and $\nu(\epsilon) \propto \epsilon^{d/2-1}$, giving $\Gamma(\epsilon) \propto \epsilon^{d/2}$.  Hence if the relaxation time is taken to be an energy-independent constant, then $\sigma_x(\epsilon) \propto \Gamma(\epsilon)$, and $S_x = -\mathcal S/N_ee$.  For non-interacting holes (\textit{i.e.} with an inverted parabolic dispersion), a similar analysis leads to $S_x = \mathcal S/N_h e$. 
 
The second limit is at temperatures much larger than the bandwidth.  This limit is most easily understood by rewriting (\ref{eq:Afunc}) as  
\begin{equation}\label{eq:Afuncmod}
F[g(\epsilon)] = \frac{\displaystyle{\int d\epsilon g(\epsilon) \Big(\frac{\epsilon}{T}\Big)\Big(-\parone{f}{\epsilon}\Big)}}{\displaystyle{\int d\epsilon g(\epsilon) \Big(-\parone{f}{\epsilon}\Big)}} - \frac{\mu}{T}.
\end{equation}
For the functions $\nu(\epsilon)$ and $\sigma_x(\epsilon)$, the upper limit in the integral is the bandwidth, and so for sufficiently large $T$ the first term tends to zero.  Since the chemical potential is \textit{defined} by $\mu/T = -(\textparone{\mathcal S}{N_e})_{E,V}$, we recover Heike's formula:
\begin{equation}\label{eq:Heikes}
-eS_x, \ \Big (\parone{\mathcal S}{N_e}\Big )_{T,V} \stackrel{\longrightarrow}{_{T\rightarrow \infty}} \Big (\parone{\mathcal S}{N_e}\Big )_{E,V} = k_B \ln \Big( \frac{2-x}{x}\Big),
\end{equation}
where $x = N_e/N_{\mathrm{cell}}$ is the filling fraction.  This result is not restricted to non-interacting systems, and can be shown to hold in general (for a system with no orbital degrees of freedom) from an analysis of the Kubo formula \cite{Chaikin}.  Note that the same analysis does not apply to $\mathcal S/N_e$, since the function $\Gamma(\epsilon)$ is non-zero outside the bandwidth, and hence the first term in (\ref{eq:Afuncmod}) is finite in the large $T$ limit.

We now consider a particular band-structure away from the two limits discussed above.  We take the 2d tight-binding model with an energy-independent relaxation time as a simple, yet non-trivial example, for which $\epsilon(\mathbf k) = -2t(\cos k_x a + \cos k_y a)$.  The relaxation time is also taken to be energy-independent: this crude approximation is sufficient for our current purpose.   

\begin{figure}[t]
 \begin{center}
  \includegraphics[width=0.45\textwidth]{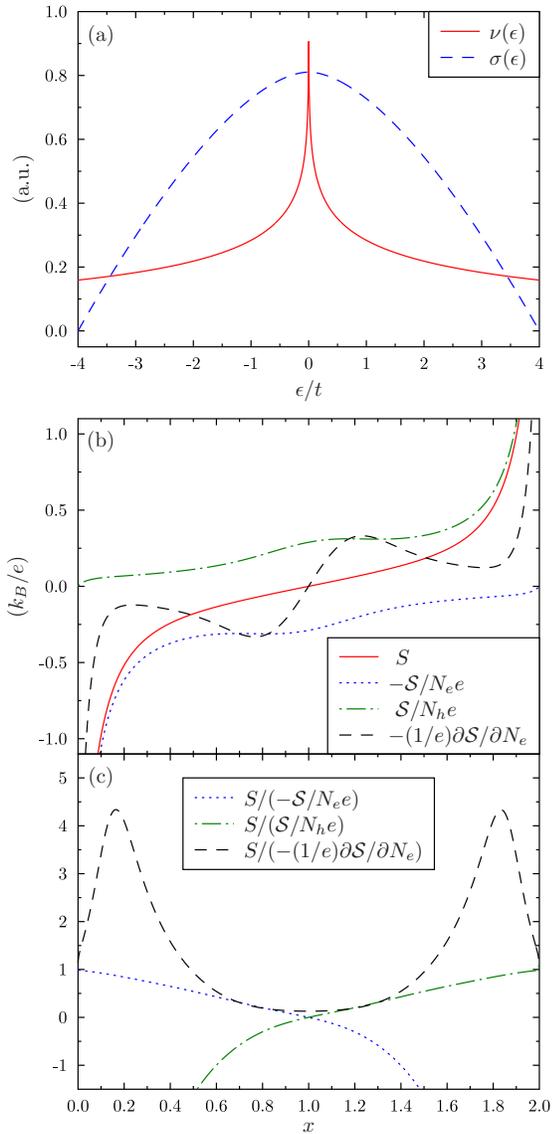}
 \end{center}
 \caption{\label{fig:tb}Thermodynamic and transport quantities calculated for the 2d tight-binding model with an energy-independent relaxation time.  (a) Density of states and energy-dependent conductivity.  (b) Thermopower, entropy per electron, entropy per hole and derivative of entropy with respect to the number of electrons, plotted as a function of filling at fixed temperature.  (c)  Ratios of the quantities presented in (b).
 }
\end{figure}    
Fig.~\ref{fig:tb}(a) shows the density of states and energy-dependent conductivity for this model.  In Fig.~\ref{fig:tb}(b) we plot the thermopower, entropy per electron $-\mathcal S/N_ee$, entropy per hole $\mathcal S/N_he$ and the quantity $-(1/e)(\textparone{\mathcal S}{N_e})_{T,V}$ at a fixed temperature $T = 0.2t$.  The thermopower is negative for $x<1$ where electrons are the dominant carriers, and positive for $x>1$ where holes dominate.  At half filling the system is particle-hole symmetric, and $S=0$.  At low filling where the dominant carriers are electrons, the thermopower tends to the entropy per electron $-\mathcal S/N_e e$.  This is shown more clearly in Fig.~\ref{fig:tb}(c), and follows from the fact that the dispersion relation is parabolic close to $\mathbf k=(0,0)$.  At large filling, the dominant carriers are holes with an inverted parabolic dispersion, and the thermopower tends to the entropy per hole, $\mathcal S/N_h e$.  Neither $-\mathcal S/N_ee$ nor $\mathcal S/N_he$ are good approximations to the thermopower close to half-filling.

In contrast, $-(1/e)(\textparone{\mathcal S}{N_e})_{T,V}$ has a qualitatively similar doping dependence to the thermopower, in the sense that it is negative for $x<1$, positive for $x>1$ and passes through zero at half-filling where the system is particle-hole symmetric.  The difference between $S$ and $-(1/e)(\textparone{\mathcal S}{N_e})_{T,V}$ arises because each $k$-state in the energy-dependent conductivity is weighted by the factor $v_x(\mathbf k)^2\tau(\epsilon(\mathbf k))$.  This weighting factor is absent in the density of states.  In essence it is this factor which converts entropy into `transported entropy', and which converts the number of electrons into the number of `carriers'.  Were it not for this factor, we would have $S=-(1/e)(\textparone{\mathcal S}{N_e})_{T,V}$ for any band-structure, and for any temperature and filling.  In the present example, this factor eliminates the strong features near the band edges and center which are present in the density of states, and which lead to the extra structure seen in $-(1/e)(\textparone{\mathcal S}{N_e})_{T,V}$.  

The advantage of $-(1/e)(\textparone{\mathcal S}{N_e})_{T,V}$ compared to the entropy per electron/hole emerges when we attempt to extend this analysis to interacting systems.  Although the general expression for the entropy in an interacting system is extremely complicated \cite{Carneiro, AGD}, a relatively simple expression may be derived for $(\textparone{\mathcal S}{N_e})_{T,V}$.  This quantity may be calculated by starting with the expression for the total number of electrons: 
\begin{equation}\label{eq:number}
N_e = 2\int d\epsilon \sum_{\mathbf k} A(\mathbf k,\epsilon)f(\epsilon). 
\end{equation}
The derivative of this expression with respect to temperature at fixed particle number vanishes, since the total number of particles is conserved.  Taking the derivative, and using the relation $(\textparone{\mathcal S}{N_e})_{T,V} = -(\textparone{\mu}{T})_{N,V}$ \cite{LL1}, we find 
\begin{equation}\label{eq:dmudt}
\Big(\parone{\mathcal S}{N_e}\Big)_{T,V}  = \frac{{\displaystyle\sum_{\mathbf k} \int d \epsilon 
\Big[ f\Big(\frac{\partial A}{\partial T}\Big)_{\mu}-A\frac{\epsilon-\mu}{T} \Big(-\frac{\partial f}{\partial 
\epsilon}\Big)\Big]}}{\displaystyle{\sum_{\mathbf k} \int d \epsilon \Big[f\Big(\frac{\partial A}{\partial \mu}\Big)_{T}-A \Big(-\frac{\partial f}{\partial \epsilon}\Big)\Big]}}.
\end{equation}
The first terms in the numerator and denominator come from the intrinsic temperature and chemical potential dependence of the spectral function.  Note that the kinetic part of the spectral function does not contain the chemical potential, as this has been shifted into the Fermi function. Hence all temperature and chemical potential dependence comes from the self-energy.

In summary, we have seen that even in a very simple example, the 2d tight-binding model with an energy-independent relaxation time, there is no universal relationship between $S$ and $-\mathcal S/N_e e$, $\mathcal S/N_h e$ or $-(1/e)(\textparone{\mathcal S}{N_e})_{T,V}$.  Exact relations only exist in particular limits.  We argued that $(\textparone{\mathcal S}{N_e})_{T,V}$ may be the more appropriate quantity to compare with the thermopower, since both reflect the particle-hole asymmetry of the system.  In a non-interacting system, the difference between these two quantities arises because $(\textparone{\mathcal S}{N_e})_{T,V}$ measures the relative asymmetry of the density of states, while the thermopower measures the relative asymmetry of the energy-dependent conductivity.  A comparatively simple expression for $(\textparone{\mathcal S}{N_e})_{T,V}$ may also be obtained in an interacting system.   

In the next section, we will first make an experimental comparison between the $c$-axis thermopower of Bi2212 and the entropy per particle obtained from specific heat measurements, and then use Eq.~(\ref{eq:dmudt}) to make a theoretical comparison with $-(1/e)(\textparone{\mathcal S}{N_e})_{T,V}$.

\section{Relating the c-axis Thermopower to the Entropy}\label{sec:caxis}

Before we compare the $c$-axis thermopower with the entropy, we first offer an argument as to why one might expect a close relationship between the two.  In the previous section, we compared the entropy for non-interacting electrons with the thermopower calculated from the Boltzmann equation in the relaxation time approximation.  We argued that one of the reasons why the thermopower and the entropy are different is that the relaxation time only enters the thermopower.  This assertion is not quite correct.  The scattering mechanism (impurities, phonons, electron-electron \textit{etc}) gives rise to a self-energy, and will in general change the equilibrium properties of the system as well.  One might imagine that if this self-energy were included in the calculation of the entropy, it would yield a closer relationship to the thermopower.  In general this argument fails, since vertex corrections mean that the transport scattering rate is not the same as the imaginary part of the self-energy.  However, this is not necessarily the case for $c$-axis transport, since we have argued in Sec.~\ref{sec:Theory} that vertex corrections may be neglected.  Hence, $c$-axis thermopower provides an `optimal' case for making the comparison between thermopower and entropy. 
\begin{figure}[t]
 \begin{center}
  \includegraphics[width=8cm,clip]{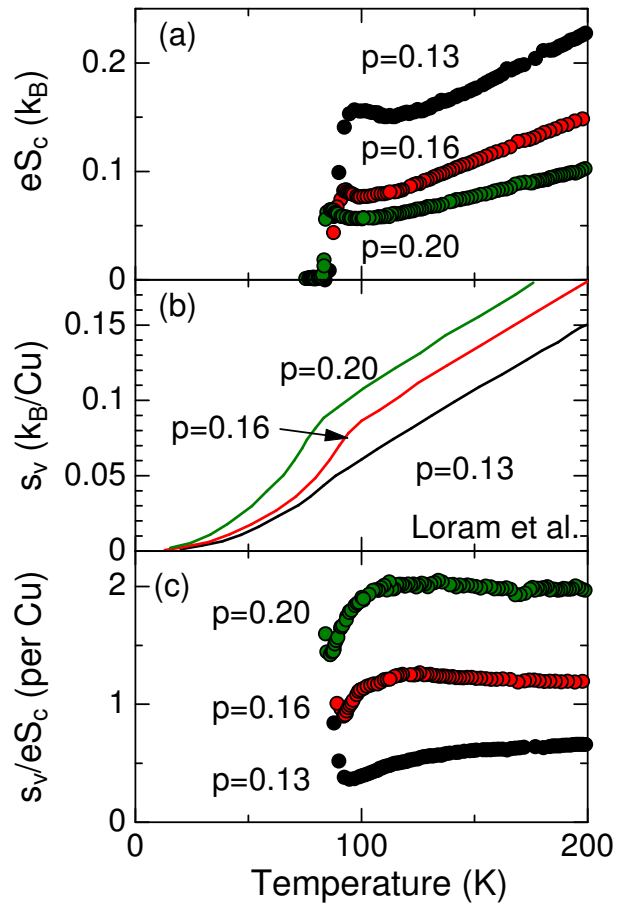}
 \end{center}
 \caption{\label{fig:exp}
 (a) $c$-axis thermopower $S_c$ of Bi2212 single crystals, reproduced from Fig.~\ref{fig:fits}.
 The data are plotted as $eS_c$ in units of $k_B$, to allow direct comparison with the entropy.
 (b) electron entropy per Cu ($s_v$) of Bi2212 single crystals
 measured by Loram \textit{et al.} plotted in a unit of $k_B$ per Cu.
 (after Ref.~\cite{loram}).
 (c) $s_v/eS_c$ calculated from (a) and (b).
 }
\end{figure}

We now compare the experimentally measured thermopower with the equilibrium entropy obtained from specific heat measurements.  For convenience, we reproduce in Fig.~\ref{fig:exp}(a) the thermopower data of Fig.~\ref{fig:fits} from 0 to 200K.  Fig.~\ref{fig:exp}(b) shows the electron entropy per Cu ($s_v$) reported in Ref.~\cite{loram}, from which we calculate $s_v/eS_c$  in Fig.~\ref{fig:exp}(c).  The monotonic increase in the entropy with temperature is very different from the in-plane thermopower \cite{TerasakiFujii}, but is much closer to that of the $c$-axis.  Now, if the thermopower measures the entropy per charge carrier (and the relevant charge carriers are the doped holes), then the ratio $s_v/eS_c$ should be independent of temperature and equal to $p$, the number of charge carriers per Cu.  Fig.~\ref{fig:exp}(c) shows that the ratio is nearly independent of temperature above around 100K, although deviates strongly as the transition is approached.  Also, although the ratio is clearly not equal to $p$, it does increase with increasing doping.  The experimental results indicate that while the $c$-axis thermopower is certainly more closely related to the entropy than is the in-plane thermopower, there is at best only a qualitative correspondence between the two.

We now turn to the theoretical comparison of the thermopower and $\eent$. Eq.~(\ref{eq:dmudt}) bears a strong resemblance to the $c$-axis thermopower (\ref{eq:caxisthermo}), although there are clearly some differences.  The tunneling matrix element appears in the thermopower but not in $\ent$.  This was to be expected, as we saw a similar difference in Sec.~\ref{sec:Results} where the group velocity entered the thermopower but not $\ent$.  Since the tunneling matrix element depends strongly on the in-plane momentum, the dominant contributions to the thermopower and $\ent$ may come from different regions of $k$-space.   

The first terms in the numerator and denominator of $\ent$ come from the intrinsic temperature and chemical potential dependence of the spectral function, and do not appear in the thermopower.  At high temperatures, we have argued that the spectral function may become rather temperature and doping independent, and hence these terms could be neglected.  The same is not true at low temperatures where the spectral function is strongly temperature dependent and very sensitive to doping.  

The second terms in the numerator and denominator of $\ent$ closely resemble their counterparts in the thermopower, except that the spectral function appears linearly rather than quadratically.  Note that if $A(\mathbf k,\epsilon) \propto A^2(\mathbf k,\epsilon)$, then because $\ent$ is a quotient it will be unchanged under the replacement $A(\mathbf k,\epsilon) \rightarrow A^2(\mathbf k,\epsilon)$. If the spectral function has any strong features, we may expect a significant difference between its first and second powers.  For a spectral function which is rather broad and featureless, this difference will be less pronounced.

For the particular case of Bi2212, the conclusion is that at high temperatures the spectral function is rather featureless, and the self-energy temperature and doping-independent.  Hence, we may expect the thermopower and $-(1/e)\ent$ to have similar temperature dependence, although the coefficient of proportionality between the two will not necessarily have any physical significance.  At lower temperatures where the spectral function does depend strongly on temperature and doping, we would not expect any kind of simple relationship between the thermopower and $-(1/e)\ent$.  Finally, it is interesting to note that the relationship we have predicted between the $c$-axis thermopower and $-(1/e)\ent$ is essentially identical to that between the thermopower and the entropy per hole seen in Fig.~\ref{fig:exp}.
 
\section{Summary}\label{sec:Summary}
In this paper we have investigated the $c$-axis thermopower in a quasi-two dimensional material.  We have shown that within the tunneling Hamiltonian formalism the $c$-axis thermopower may always be written as a Mott formula in terms of the energy-dependent conductivity, $\sigma_c(\epsilon) = 2e^2\sum_{\mathbf k}t_{\mathbf k}^2 A^2(\mathbf k, \epsilon)$.  This implies that the $c$-axis thermopower is a measure of the relative asymmetry of the spectral function around the Fermi energy. 

Using this theory, we have shown that the $c$-axis thermopower and resistivity imply that the symmetric and antisymmetric parts of the spectral function become approximately temperature independent at high temperatures.  At lower temperatures, we have suggested that the up-turn in the $c$-axis thermopower may reflect the opening of the pseudogap.  It is also possible that very close to $T_c$ superconducting fluctuations may also contribute to this upturn.  The main conclusion of our analysis is that the $c$-axis thermopower is an effective probe of the particle-hole asymmetry in the system.  We believe that a systematic analysis of $c$-axis resistivity and thermopower data could provide valuable insights about the spectral function, particularly in the pseudogap regime.  The $c$-axis resistivity and thermopower data together would also form a stringent test of any spectral function obtained phenomenologically, or from a microscopic model.

The second theme of the paper is the relationship between thermopower and entropy.  We discussed this relationship in general terms, and suggested that the derivative of the entropy with respect to particle number may be more suitable to compare with the thermopower, rather than the entropy per particle.  We argued that the $c$-axis thermopower of a quasi-two dimensional system is a case where there might exist a close relationship with the equilibrium entropy.  In the case of Bi2212, we showed experimentally that, at best, there is only a qualitative correspondence between the $c$-axis thermopower and the entropy per hole at moderate temperatures.  Having derived expressions for both the $c$-axis thermopower and $\ent$, we were able to make an explicit comparison between the two.  Our analysis indicated that the relationship between these two quantities would be similar to that between the thermopower and entropy per hole seen experimentally.  Because of the existence of a relatively simple expression for $\ent$, even in an interacting system, we believe that this quantity will prove useful in investigating the relationship between thermopower and entropy in other systems.

\acknowledgments
The authors would like to thank K. Tanaka,
A. Fujimori and T. Tohyama for fruitful discussion.
This research was partially supported by the Ministry of Education, 
Culture, Sports, Science and Technology, 
Grant-in-Aid for Scientific Research (C),
2001, No. 13640374.  We also gratefully acknowledge the support of The Royal Society and EPSRC.

\end{document}